\documentstyle[12pt]{article}
\begin{document}

\begin{center}

{\large\bf Comments on} \\
{\large\bf ``Modified negative binomial description of the
multiplicity distributions in lepton-nucleon scattering''}

\vspace{.5cm}

{\sc S. Hegyi}\footnote{E-mail: hegyi@rmki.kfki.hu}

\vspace{.2cm}

{\normalsize
	Research Institute for
	Particle and Nuclear Physics, \\
	H--1525 Budapest 114, P.O. Box 49.
	Hungary
}

\end{center}

\pagestyle{plain}
\noindent
In two recent papers the present author introduced a generalization of the
Negative Binomial Distribution~[1,2]. It was obtained by extending the
validity of the asymptotic KNO scaling form of the NBD to the $\mu>0$
powers of the scaling variable $x=n/\bar n$. With this modification the
asymptotic KNO function becomes
\begin{equation}
        f(x)=\frac{\mu}{\Gamma(k)}\,\lambda^{\mu k}
        x^{\mu k-1}\exp\left(-[\lambda x]^{\mu}\right)\label{1}
\end{equation}
which is the generalized gamma distribution. In pQCD
$\mu=(1-\gamma)^{-1}$ with $\gamma\propto\sqrt{\alpha_s}$
being the QCD multiplicity anomalous dimension. Poisson transforming
$f(x)$ one obtains the cited generalization of the NBD whose analytic
form is given by
\begin{equation}
   P_n=
   \frac{1}{n!\,\Gamma(k)}\;{\sf H}^{1,1}_{1,1}
   \left[
      \,\frac{1}{\theta}\left|
      \begin{array}{c}
         (1-n,\; 1)             \\
         (k,\; 1/\mu)
      \end{array}
   \right]\right.\quad\mbox{for}\quad0<\mu<1
\end{equation}
and
\begin{equation}
   P_n=
   \frac{1}{n!\,\Gamma(k)}\;{\sf H}^{1,1}_{1,1}
   \left[
      \,\theta\left|
      \begin{array}{c}
         (1-k,\; 1/\mu)         \\
         (n,\; 1)
      \end{array}
   \right]\right.\quad\mbox{for}\quad\mu>1
\end{equation}
where $\theta=\bar n\Gamma(k)\Gamma^{-1}(k+1/\mu)$ and ${\sf H}(\cdot)$
is the Fox generalized hypergeometric function or ${\sf H}$-function.
For $\mu=1$ the NBD is recovered. For $\mu>1$ $P_n$ is not infinitely
divisible and the factorial cumulants of the distribution exhibit
nontrivial sign-changing oscillations. Eqs.~(2)-(3) involve as special
and limiting cases the Poisson transform of numerous classical probability
densities such as the Chi, Weibull, Maxwell, Pareto and Lognormal to
mention but a few. Since the name Generalized NBD is already in use (not
rarely to denote different discrete distributions) we shall call
the above ${\sf H}$-function extension of the Negative Binomial
as HNBD for short.

In ref.~[2] the HNBD was used to analyse the charged particle multiplicity
distributions measured by the H1 Collaboration in deep inelastic $e^+p$
scattering at HERA~[3]. Similar analysis was performed in ref.~[4] using
another 3-parameter generalization of the NBD, to so-called Modified NBD.
The generating function $G(z)=\sum_{n=0}^\infty P_nz^n$ of the MNBD is
given by
\begin{equation}
	G(z)=\left(\frac{1+\Delta(1-z)}{1+r(1-z)}\right)^k
	\quad\mbox{with}\quad r=\Delta+\bar n/k
\end{equation}
which yields for $\Delta=0$ the generating function of the NBD.
The analytic form of the MNBD is usually written as
\begin{eqnarray}
	P_0&=&\left(\frac{1+\Delta}{1+r}\right)^k \nonumber \\
	P_{n\geq1}&=&\frac{1}{n!}\left(\frac{\Delta}{r}\right)^k
	\left(\frac{r}{1+r}\right)^n\sum_{j=1}^k{}_kC_j
	\frac{\Gamma(n+j)}{\Gamma(j)}
	\left(\frac{r-\Delta}{(1+r)\Delta}\right)^j.
\end{eqnarray}
During the past few years the MNBD was successfully used to describe the
multiplicity distributions measured in different collision processes.
For the H1 data the quality of the MNBD fits reported in~[4]
are again very satisfactory.
The author of ref.~[4] made a comparison of the MNBD and HNBD
results. It was found that the $\chi^2/\mbox{d.o.f.}$ values
obtained by the one-parameter
MNBD fits are significantly smaller than those obtained
by the two-parameter HNBD fits.
This seems to indicate that the HNBD is less satisfactory in
reproducing the H1 data than the MNBD. But the reason behind the better
quality of MNBD fits is different.

In ref.~[2] the HNBD was fitted to the corrected
H1 data with variable parameters
$\bar n$ and $\mu$. The shape parameter was fixed at $k=1$
(Weibull case). As is clearly stated in [2] the usually less
than 5\% relative errors of the best-fit parameters
are only statistical, the systematic uncertainties
of the H1 data were not taken into account in the analysis.
The MNBD was examined in ref.~[4] with
fixed shape parameter $k=7$ and variable $\Delta$.
The value of $\bar n$
was taken from ref.~[3]. Although it is not stated explicitly,
the 15--25\% relative errors of the single fit parameter $\Delta$
indicate that the MNBD analysis was carried out
using both the statistical and
systematic uncertainties of the H1 data. The comparison of
the quality of fits obtained in refs.~[2] and~[4]
is therefore misleading.

To clarify the situation the present author repeated the HNBD
analysis  with the inclusion of both the statistical
and systematic errors of
the H1 data. Following ref.~[4] $\bar n$ was fixed at its experimental
value and only $\mu$ was taken as free parameter
($k=1$ as earlier).
The comparison of the MNBD and HNBD fits is shown in
Table~1. We can see that the HNBD easily reproduces the quality
of MNBD fits. Of course it is not particularly meaningful to
speculate which distribution is better; neither of them can be ruled
out by a $\chi^2$-analysis of the H1 data tabulated in~[3].
As a more stringent test, it
would be of interest to compare MNBD and HNBD fits with the use
of the full covariance matrix which makes possible to take into
account the correlations between adjacent multiplicities.

\vskip1cm

\begin{tabular}{||c||c|c||c|c||}\hline
$W$ (GeV) & $-\Delta$ & $\chi^2/$ d.o.f. & $\mu$ & $\chi^2/$ d.o.f.    \\
\hline\hline
\ $80\div115$ & $0.216\pm0.034$ & \ 2.7/18  & $4.811\pm0.392$ & 0.9/18 \\
\hline
 $115\div150$ & $0.233\pm0.036$ & 10.6/21 & $4.238\pm0.277$ & 2.2/21   \\
\hline
 $150\div185$ & $0.187\pm0.049$ & \ 4.6/22  & $4.011\pm0.288$ & 1.9/22 \\
\hline
 $185\div220$ & $0.264\pm0.056$ & 11.3/23 & $4.021\pm0.334$ & 1.5/23   \\
\hline
\end{tabular}

\bigskip\noindent
Table 1. \ Comparison of the MNBD fits, taken from ref.~[4], and the HNBD
fits for the H1 multiplicity data corresponding to the pseudorapidity
interval $1<\eta^*<5$. The errors quoted are statistical and systematic.

\end{document}